\documentclass[12pt]{article}

\usepackage{amsfonts}

\newif\ifpdf
\ifx\pdfoutput\undefined
\pdffalse 
\else
\pdfoutput=1 
\pdftrue
\fi

\usepackage{graphicx}

\title{Quantum Watermarking by Frequency of Error when Observing Qubits in Dissimilar Bases}
\author{G Gordon Worley III\\University of Central Florida\thanks{Thanks to Dan C. Marinescu for his advice on the completion of this paper.}\\ge011023@pegasus.cc.ucf.edu}

\begin{document}


\maketitle

\abstract{We present a so-called fuzzy watermarking scheme based on the relative frequency of error in observing qubits in a dissimilar basis from the one in which they were written.  Then we discuss possible attacks on the system and speculate on how to implement this watermarking scheme for particular kinds of messages (images, formated text, etc.).}

\section{Introduction}

Alice wrote a digital message that she wants to send to her friend Bob.\footnote{Alice and Bob are standard characters used in illustrations of message exchange systems.}  And although she trusts Bob enough to send him this message, she wants to be sure that he will not try to claim its authorship.  To accomplish this, Alice will insert a watermark in her message that has the following properties\cite{informationhiding}:

\begin{enumerate}
\item
It is not perceptible.  Bob will not notice the presence of the watermark.
\item
It is nondestructable, unless the message is destroyed beyond use in the process.
\item
It is uniquely verifiable by a secret that only Alice has.
\end{enumerate}

Alice will accomplish this using a watermarking scheme.  In the last decade, many researchers have proposed schemes of varying quality using classical information theory.\cite{informationhiding}  Quantum information theory, however, has been ignored.  Yet it offers capabilities that do not exist in classical information theory alone.\cite{fundqit}  Combined with the evidence of advantages quantum information theory has conveyed to other coding problems\cite{qcintro, qufinger, qudigsig, sealingqumsg}, we believe that it can improve watermarking schemes, although to the best of our knowledge no one has yet published on this topic.

Here we present a watermarking scheme based on the relative frequency of error in observing qubits in a dissimilar basis from the one in which they were written.

\section{A Quantum Watermarking Scheme}

Alice has some quantum message $M$ that she wants to watermark and changes it to $\tilde{M}$, which contains some watermarking qubits.  All qubits $|\phi_i\rangle \in M$ are written in some basis $j$.  To watermark $M$, Alice observes the qubits $|\phi_i\rangle$ where $i \in \mathcal{I}$, where $\mathcal{I}$ is the set of bits in $M$ that we will use for watermarking, and writes them back to $M$ in a dissimilar basis $k$, producing $\tilde{M}$, the watermarked message.  $\tilde{M}$ now contains the original qubits written in basis $j$ and the watermarking qubits at $|\tilde{\phi}_i\rangle \in \tilde{M}$ where $i \in \mathcal{I}$ written in basis $k$, where $k \ne j$.

The watermark is created when $\tilde{M}$ is observed in the basis $j$ (we denote this as $\tilde{M} \circ j$).  The watermarking qubits, which are written in basis $k$, will be observed in error of the intended value with a probability of error $p_e$.  The watermark is the relative frequency of error in the bits $a_i \in \tilde{M} \circ j$ where $i \in \mathcal{I}$, where $\mathcal{I}$ and $k$ are the secrets.

\begin{figure}[htb]
\center\includegraphics[bb=0pt 0pt 256pt 240pt]{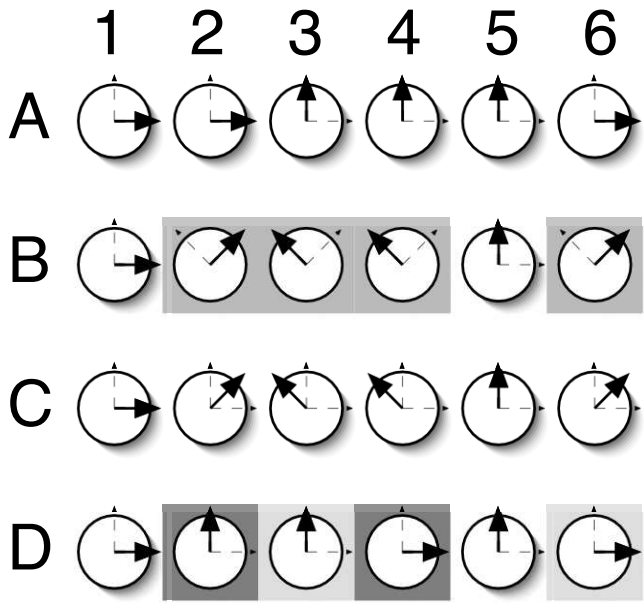}
\caption[encoding]{\label{encoding} The watermarking process}
\end{figure}

Consider the example in Figure \ref{encoding}, where circles represent qubits, dashed lines represent basis states, and solid lines represent the current state of a qubit.  We begin with our message encoded as qubits in {\bf A}.  Then in {\bf B} the greyed qubits (i.e. $\mathcal{I} = {2, 3, 4, 6}$) are rewritten in a different, secret basis by Alice.  She then sends {\bf B} to Bob, who interprets the message as in {\bf C} because he does not know which qubits are written in a different basis.  When Bob observes the message, he gets {\bf D} with some of the watermarking qubits {\it in error} (the dark grey qubits) and some {\it accurately} (the light grey qubits), producing a fuzzy watermark (see Figure \ref{verification} for an example of verification).

\subsection{Verification}

One day Alice visits Bob's Website and finds a message $\tilde{M}'$ that she believes to be a copy of $M$.  Since she watermarked $M$ before sending it to Bob, she can detect if $\tilde{M}'$ is her watermarked $M$ by comparing the bits that result from observing $\tilde{a}_i' \in \tilde{M}' \circ j$ and $a_i \in M \circ j$, where $i \in \mathcal{I}$.  If the relative frequency of error between $\tilde{a}_i'$ and $a_i$ is very nearly the expected probability of error of reading a qubit as if it were written in basis $j$ when it was actually written in basis $k$, the message $\tilde{M}'$ is probably $M$ watermarked by Alice.

We say `probably' because the strength of the system depends on the probabilities associated with observing qubits.  For example, a qubit written in a certain basis might have a value $|\phi\rangle = \alpha_0|45^\circ\rangle + \alpha_1|135^\circ\rangle$.  When $|\phi\rangle$ is read in a different basis from the one it was written, say the $\alpha_0|0^\circ\rangle + \alpha_1|90^\circ\rangle$ basis, the value of the bit observed will be in error at a certain rate.  If $|\phi\rangle = 1|45^\circ\rangle + 0|135^\circ\rangle$, then $|\phi\rangle = \sqrt{0.5}|0^\circ\rangle + \sqrt{0.5}|90^\circ\rangle$, so the probability of error when read in the second basis is 0.5 from the expected value when read in the first basis.  When given enough of these qubits written in one basis and read in another, we can expect the relative frequency of error to approach the expected probability of error.  Specifically, we require at minimum $n$ bits, where $n = a + b$ s.t. $\frac{a}{a + b} \geq p_e$ bits for $0 \leq p_e \leq 1$.

\begin{figure}[htb]
\center\includegraphics[bb=0pt 0pt 272pt 175pt]{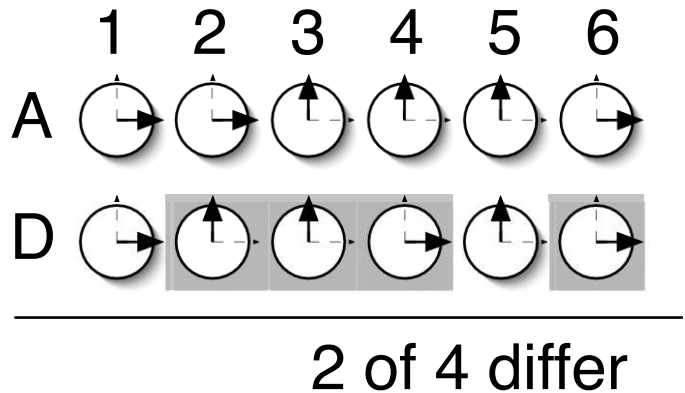}
\caption[verification]{\label{verification} The watermark verification process}
\end{figure}

Returning to the example from Figure \ref{encoding}, Figure \ref{verification} shows how Alice would verify that {\bf D} is the same message as {\bf A} with her watermark.  Here Alice compares the two messages on the qubits in $\mathcal{I}$ and finds the relative frequency of error.  In this case, there are 2 errors in 4 bits, for a relative frequency of 0.5.  Since we can assume from Figure \ref{encoding} that $p_e = 0.5$, Alice has verified that {\bf D} is a watermarked {\bf A}.

Clearly this example's message is far too short to be watermarked in the real world, but it demonstrates the system.  We discuss real world use in the next section.

\subsection{Discussion}

The system is intended to be used in the following way.  Alice has a message made of qubits or a classical message that she converts to qubits.  After inserting the watermarking qubits, she can either observe the message immediately and only send the classical version out, or leave the message in qubit form and let Bob observe the qubits.  Note that the watermark is not actually created until observation, so it is probably best for Alice to only send out a pre-observed version of the watermarked message to avoid the averaging attack (see {\bf Attacks} below).

$\mathcal{I}$ should be chosen such that if $\tilde{a}_i \ne a_i$, the message is not perceptibly changed, and such that the watermark cannot be easily removed by non-destructive manipulations.  Also, $\mathcal{I}$ can be chosen according to some secret key $K$ such that only the holder of $K$ will be able to find $\mathcal{I}$ within a reasonable amount of time, although in reality $\mathcal{I}$ is the secret that must be kept; $K$ is a subsidiary secret that can be used for convenience given a reliable function $f: K \rightarrow \mathcal{I}$.

The key to the effectiveness of this system is in the choice of $\mathcal{I}$ and the value of $p_e$.  The errors the system will introduce must be imperceptible, else an attacker may be able to locate enough of $\mathcal{I}$ to disable the watermark without destroying $M$.  Also, the size of $\mathcal{I}$ is important, because if it is too small the relative frequency of error between $\tilde{a}'_i$ and $a_i$ will not reliably converge on $p_e$.  As stated earlier, $|\mathcal{I}| \geq n$, $n = a + b$ s.t. $\frac{a}{a + b} \geq p_e$, but in general $|\mathcal{I}|$ should be much greater than $n$ to improve the reliability and robustness of the watermark (the exact number will vary with message format and how close the relative frequency of error must come to the expected probability of error).

Obviously, this scheme is equivalent to a classical scheme that flips the bits $a_i \in M$ for $i \in \mathcal{I}$ with a certain probability $p_e$.  The advantage of the quantum version is that, unlike the classical, it does not rely on a pseudo random number generator, or PRNG.  PRNGs need to collect entropy from many low-entropy sources since high-entropy sources, like radioactive decay, are generally unavailable, thus the quality of random numbers may suffer.\cite{randomnumbergeneration}  Further, as Kelsey et al. have shown \cite{prngsattacks}, most PRNG do not produce good random numbers, making it easier to crack implementations of coding schemes that use them.  Depending on the size and number of messages being watermarked, the scheme may suffer because there is not enough entropy to generate real random numbers.  A quantum system, though, relies on the physical randomness of observing a qubit with an ambiguous state.  We obtain real randomness from quantum observation that does not run out of entropy the way a PRNG does.  This is the key advantage over a purely classical version of this scheme.

Using qubits as RNGs depends on a nondeterministic quantum theory.  If it turns out that the observation of qubits is determined by hidden variables or other factors, this scheme (and others that depend on qubits for randomness) may break.

\section{Attacks}

Alice should be careful not to release multiple watermarkings of the message (that is to say, more than one $\tilde{M}$ for a particular $M$).  Bob could use these multiple versions to average the watermarked messages to approximate or even find the original, unwatermarked, message.  This is possible because for each $\tilde{a}_i' \in \tilde{M}' \circ j$ where $i \in \mathcal{I}$ there is only a difference from $a_i$ with a probability of $p_e$.  If he can get enough versions of $\tilde{M}$, he can find most if not all of $\mathcal{I}$ by looking at where the bits differ and, guessing $k$, average the values of the watermarking bits to produce a version of the message without a valid watermark but imperceptibly different from the original.  This attack might be made harder if $\mathcal{I}$ varies for each $\tilde{M}$ for a particular $M$, however it would not completely eliminate the possibility of successful averaging.

If $|\mathcal{I}|$ is too small, a sufficient but imperceptible amount of noise will reduce the number of intact $\tilde{a}_i' \in \tilde{M}' \circ j$ such that the relative frequency of error will not reliably approach the expected probability of error.  This happens because the scheme depends on the frequency with which $a_i$ disagrees with $\tilde{a}_i$, so if some of $\tilde{M}'_i$ are changed, the relative frequency of error may no longer approach $p_e$.  The best defense against this is to increase $|\mathcal{I}|$, but not to the point that the watermark becomes perceptible.  It should be that the noise needed to disable the watermark will render the message useless.

Attacks that do not damage the message but change the indices of bits by padding the message with nondestructive bits are particularly damaging.  For example, depending on the format of the message, it may be that the bits $\tilde{a}_0, \tilde{a}_1, \tilde{a}_2, \dots, \tilde{a}_n$ for $\tilde{a}_i \in \tilde{M}$ can be shifted $+1$, so that $\tilde{a}_i$ becomes $\tilde{a}_{i+1}$ and $\tilde{a}_0$ receives a value that is imperceptible or ignored by the format.  Defeating this attack is nontrivial, because it requires recognizing how the bits in the message have been transformed and reversing the transformation so that the indices $a_i$ and $\tilde{a}'_i$ match up.

\section{Implementation}

We have not said anything regarding how $\mathcal{I}$ should be chosen.  This is particular to the kind of message (image, formatted text, audio, etc.) and the required robustness, although some formats are more likely to accommodate this fuzzy watermarking scheme.  Images and audio, for example, have room for imperceptible noise.  Video, in some cases, has enough room to allow steganographic audio to be transmitted, which provides more than enough room for a fuzzy watermark.\cite{informationhiding}  These message types provide plenty of room for our watermarking scheme.  Conversely, text, even formatted text, may be too fragile for this system, quickly dissolving into perceptible noise.

Ultimately, we leave it as an exercise for those who wish to implement this scheme with a particular format to decide how to choose $\mathcal{I}$.

\section{Conclusion}

Quantum watermarking is a new-born subfield of the young field of digital watermarking.  Although the first steps taken in this paper do not provide stunning results like those found in quantum cryptography, this may change in the near future.

\bibliographystyle{unsrt}
\bibliography{qmfe}

\begin{thebibliography}{1}

\bibitem{informationhiding}
Stefan Katzenbeisser and Fabien~A.P. Petitcolas, editors.
\newblock {\em Information Hiding: techniques for steganography and digital
  watermarking}.
\newblock Artech House, Norwood, MA, 2000.

\bibitem{fundqit}
Michael Keyl.
\newblock Fundamentals of quantum information theory, February 2002.
\newblock arXiv:quant-ph/0202122.

\bibitem{qcintro}
Eleanor Rieffel and Wolfgang Polak.
\newblock An introduction to quantum computing for non-physicists, January
  2000.
\newblock arXiv:quant-ph/9809016.

\bibitem{qufinger}
Harry~Buhrman et~al.
\newblock Quantum fingerprinting, February 2001.
\newblock arXiv:quant-ph/0102001.

\bibitem{qudigsig}
Daniel Gottesman and Isaac~L. Chuang.
\newblock Quantum digital signatures, November 2001.
\newblock arXiv:quant-ph/0105032.

\bibitem{sealingqumsg}
H.~F. Chau.
\newblock Sealing quantum message by quantum code, August 2003.
\newblock arXiv:quant-ph/0308146.

\bibitem{randomnumbergeneration}
Chapter 6: Random number generation.
\newblock Web PDF.
\newblock Accessed on 27 June, 2001 from
  http://www.cypherpunks.to/~peter/06\_random.pdf. This appears to be a book
  chapter, but I cannot discern what book.

\bibitem{prngsattacks}
John~Kelsey et~al.
\newblock Cryptanalytic attacks on pseudorandom number generators.
\newblock In {\em Fast Software Encryption, Fifth International Workshop
  Proceedings}, pages 168--188. Springer-Verlag, March 1998.
\newblock Accessed on 19 December, 2003 from
  http://www.schneier.com/paper-prngs.pdf.

\end{thebibliography}

 \end{document}
 \end